# Preferential Partner Selection in an Evolutionary Study of Prisoner's Dilemma

**Dan Ashlock[†], Mark D. Smucker[§], E. Ann Stanley[†‡], and Leigh Tesfatsion[*†]**

[†] Department of Mathematics, 400 Carver, and [*] Department of Economics, 375 Heady Hall, Iowa State University, Ames, Iowa 50011-1070,

[§] Department of Computer Sciences, 1210 West Dayton Street, University of Wisconsin-Madison, Madison, Wisconsin 53706.

[‡] Proofs and inquiries should be sent to this author. email: stanley@iastate.edu, FAX: (515)294-5454, Office phone: (515)294-7965.

## Abstract

Partner selection is an important process in many social interactions, permitting individuals to decrease the risks associated with cooperation. In large populations, defectors may escape punishment by roving from partner to partner, but defectors in smaller populations risk social isolation. We investigate these possibilities for an evolutionary prisoner's dilemma in which agents use expected payoffs to choose and refuse partners. In comparison to random or round-robin partner matching, we find that the average payoffs attained with preferential partner selection tend to be more narrowly confined to a few isolated payoff regions. Most ecologies evolve to essentially full cooperative behavior, but when agents are intolerant of defections, or when the costs of refusal and social isolation are small, we also see the emergence of wallflower ecologies in which all agents are socially isolated. In between these two extremes, we see the emergence of ecologies whose agents tend to engage in a small number of defections followed by cooperation thereafter. The latter ecologies exhibit a plethora of interesting social interaction patterns.

# 1 Introduction

Following the path-breaking work of Axelrod (1984; 1987; 1988), the Iterated Prisoner's Dilemma (IPD) is now commonly used by researchers to explore the potential emergence of mutually cooperative behavior among non-altruistic agents. See, for example, Miller (1989) and Lindgren and Nordahl (1994). These studies have shown that mutually cooperative behavior tends to emerge if the number of game iterations is either uncertain or infinite, the frequency of mutually cooperative play in initial game iterations is sufficiently high, and the perceived probability of future interactions with any given current opponent is sufficiently large.

Most studies of IPD assume that individual players have no control over which opponents they play. Players are matched as game partners either randomly or by means of a deterministic mechanism such as round-robin or grid neighborhood play. In real-life situations, however, agents are not always prisoners who have no alternative but to play their assigned PD games. Instead, social interactions are often characterized by the preferential choice and refusal of partners. In what ways, then, might the introduction of preferential partner selection change the nature of the IPD?

Previous research suggests that, depending upon the precise population structure, the decision rules used for partner selection, and the penalties imposed for rejected offers and for deciding not to play, cooperators or defectors may benefit from preferential partner selection. For example, Kitcher (1992) and Schuessler (1989) show that the option of refusing to play previously defecting players can increase the fitness of cooperative players and allow them to invade defecting populations. Orbell and Dawes (1993) argue that it is to the benefit of society as a whole to evolve social structures that allow individuals to opt out of games. Their experiments indicate that humans who are themselves cooperatively inclined tend to be more optimistic about the cooperative intentions of other players and hence to play more games. In Eshel and Cavalli-Sforza (1982), beneficial assortative mixing may occur either because agents playing the same strategy are more likely to encounter each other, or because agents playing cooperative strategies actively select each other.

The ability to actively seek out known cooperators as partners also provides an incentive for agents to be reliably cooperative, so that they will be chosen as partners, and this potentially increases the incidence of cooperation in a society (Tullock, 1985). Hirshleifer and Rasmussen (1989) find that group ostracism can permit cooperative agents to protect themselves from defectors. On the other hand, Dugatkin (1991) shows that the ability to choose partners in large populations divided into isolated patches may permit roving defectors to move from one patch to the next, avoiding ostracism while taking advantage of each patch in turn.

Finally, the introduction of preferential partner selection results in social networks of interacting players. Who chooses whom, and why, affects who does well, and this in turn affects the outcomes of the overall game. Questions about social network formation are key to understanding societies. How do groups form? What roles do highly connected individuals play? Social networks are also interesting because they are pathways for the transmission of diseases, information, and cultural traits.

In a previous paper (Stanley et al., 1994), we studied an IPD choice and refusal mechanism that combines active choice of potential game partners with the ability to refuse play with those judged to be intolerable. Players use continually updated expected payoffs to assess the relative desirability of potential partners. This use of expected payoffs is meant to capture the idea that players attempt to select partners rationally, using some degree of anticipation, even though they do not know their partners' strategies and payoffs. Our choice and refusal mechanism is thus more flexible and general than many of the mechanisms studied by previous researchers, although it does not currently allow for the information exchange between players assumed by Kitcher. Also, we considered a single small population, so that defectors cannot rove from one isolated population to the next, as in Dugatkin (1991), but instead risk eventual ostracism.

In particular, we studied how the ability both to choose and to refuse potential game partners affects interactions among a small set of simple IPD strategies, and we used a five-player population to illustrate the formation of social networks. We also conducted a number of evolutionary simulations. The interaction dynamics in both our analytical and simulation studies were seen to be complex, even for small populations. Choice is used by all players to home in quickly on those who will cooperate with them. This permits nice players to interact with each other, but also allows predatory individuals to locate and form long term relationships with victims within the limit of occasional defection tolerated by the refusal mechanism. On the other hand, refusal ensures that very nasty players do poorly, since repeatedly defecting players are typically ostracized as other players increasingly refuse their offers. Indeed, wallflower populations sometimes emerged in which all players defected until they became solitary, neither making nor accepting game offers from other players. Overall, however, we observed cooperation to emerge much more quickly and frequently with choice and refusal of partners than with round-robin matching.

In this paper we present a variety of new analytical and simulation findings on the evolutionary IPD with choice and refusal of partners, or evolutionary IPD/CR for short. We first review in Section 2 the basic structure and implementation of the evolutionary IPD/CR. In particular, we discuss the finite state machines used to represent players' IPD strategies as well as the genetic algorithm used to evolve the player population from an initial, randomly-chosen population.

Just as humans cannot instantly evolve wings, so players in our co-evolutionary framework cannot necessarily jump from a defecting mode of behavior to a cooperative one. In particular, the genetic material available in our initial population constrains its future evolution. This path dependence turns out to be particularly important for the interpretation of the IPD/CR simulation studies reported in the present paper, since we work with relatively small populations of thirty players. Section 3 thus undertakes an analytical characterization of the distribution of behaviors in the initial player population. In particular, it is shown that a uniformly distributed selection of genetic structures for these players implies a nonuniform selection of their IPD strategies, one that is highly biased towards simple strategies.

In the final two sections we detail a series of simulation studies that have been conducted to explore the sensitivity of evolutionary IPD/CR outcomes to changes in key parameters. In particular, we first describe one-parameter and two-parameter sensitivity studies for the parameters characterizing the choice and refusal mechanism. We then report on experiments conducted to test the sensitivity of outcomes to changes in the potential complexity of players' IPD strategies, as measured by the number of states in their finite state machine representations. Also, we briefly summarize preliminary studies in which two key parameters characterizing the choice and refusal mechanism are incorporated into the genetic structure of each player and allowed to evolve over time. Finally, we discuss the sensitivity of the behavioral diversity of our populations to changes in the implementation of the genetic algorithm used to evolve our player populations.

| | *Player 1* | |
|---|---|---|
| *Player 2* | *c* | *d* |
| *c* | 3 | 5 |
| *d* | 0 | 1 |

Table 1: The PD payoff matrix for player 1 used in all simulations. Player 1's moves are given across the top and player 2's down the side.

# 2 The Evolutionary IPD/CR Simulations

Each simulation discussed in this paper is initialized with a randomly generated population of $N$ players, whose genetic code specifies their strategies for playing iterated prisoner's dilemma. The simulation then consists of a sequence of generations inter-spaced with genetic steps. Each generation engages in an IPD/CR tournament consisting of $I$ iterations in which the agents choose and refuse partners for games of prisoner's dilemma. In the genetic step, the scores from this tournament are used to assess each player's relative fitness and a new population of $N$ players is then generated via a genetic algorithm with crossover and mutation.

## 2.1 Prisoner's Dilemma

The one-shot prisoner's dilemma game is a game between two players. Each player has two possible moves, cooperate $c$ or defect $d$, and each player must move without knowing the move of the other player. If both players defect, each receives a payoff $D$. If both cooperate, each receives a payoff $C$ which is strictly greater than $D$. Finally, if one defects and the other cooperates, the cooperating player receives the lowest possible payoff $L$ and the defecting player receives the highest possible payoff $H$, where $L < D < C < H$. The payoffs are also restricted to satisfy $(L + H)/2 < C$, so that unsynchronized alternations of cooperation and defection with the same partner do not yield as high an average payoff as repeated mutual cooperation. Table 1 shows the particular values of these payoffs used in our simulations, which satisfy these restrictions.

The dilemma in the one-shot PD game is that, if *both* players defect, both do worse than if both had cooperated, yet there is always an incentive for an individual player to defect. The best response to defection is to defect, because this avoids the lowest payoff, $L$; and the best response to cooperation is to defect, because this achieves the highest payoff, $H$.

## 2.2 Choice and Refusal of Partners

There are many different procedures which individuals might use to select partners. In deciding how to model partner selection, we considered situations, such as the dating game, where people interact locally and want to pair up either with their best choice or with someone who directly approaches them and offers to interact.

Specifically, we assume that there is a potential cost to making a PD game offer—such as the shame of being refused—whereas receiving offers is costless. In addition, we impose a (possibly small) cost on wallflower players who neither make nor accept offers, which is intended to deter this kind of behavior. We also assume that each individual considers the merits of each offer separately from every other offer, implying that a player with many offers may play many games.

As a simple way of ranking their potential partners, our players use expected payoffs to keep a running tab on how well they are doing against each other player. While real humans presumably use more complex rules to select partners, this method of ranking individuals allows us to begin to study the effects of partner selection from a broader perspective than previous researchers.

## 2.3 The IPD/CR Tournament

We study a slightly simplified version of the choice and refusal tournament of Stanley et al. (1994). Each generation engages in a tournament consisting of $I$ iterations. At the beginning of each iteration $i \geq 1$, each player $n$ associates an expected payoff with each other player $m$, denoted by $\pi_{i-1}(m|n)$. This expected payoff is used to determine which players are tolerable as partners and which player it will choose.

Given any player $n$, another player $m$ is *tolerable* for player $n$ in iteration $i$ if and only if

$$\pi_{i-1}(m|n) \geq \tau \ , \tag{1}$$

where $\tau$ is the *minimum tolerance level*. If any players are tolerable to player $n$, then it makes an offer to the player $m$ for whom its expected payoff $\pi_{i-1}(m|n)$ is highest, with any ties being settled by a random draw.

After these choices are completed, each player is given an opportunity to refuse offers. Each player refuses all PD game offers received from intolerable players and accepts all others. Players cannot opt out of an offer received from someone they judged to be tolerable at the beginning of the iteration. Each time an offer is refused, the player making the offer receives a *refusal payoff*, $R$. A player who finds all players intolerable, and hence who neither makes nor accepts any offers, receives a *wallflower payoff*, $W$. All accepted PD game offers are then played. Even when two players choose each other, they only play one PD game with each other.

In the initial iteration 1, prior to any interactions, all players have the same *initial expected payoff* $\pi_0$ for each player. After this, expected payoffs are modified whenever two players interact. Consider any two players $n$ and $m$. If $n$ neither made nor accepted a PD game offer from $m$ in the current iteration $i$, then $n$'s expected payoff $\pi_{i-1}(m|n)$ for the play of a PD game with $m$ in iteration $i$ is not changed. On the other hand, suppose player $n$ either received a PD payoff or a refusal payoff from interacting with player $m$. Let this payoff be denoted by $U$. Then player $n$'s expected payoff from player $m$ is modified by taking a weighted average over player $n$'s payoff history with player $m$,

$$\pi_i(m|n) \ = \ \omega\pi_{i-1}(m|n) + (1-\omega)U \ , \tag{2}$$

where the *memory weight* $\omega$ controls the relative weighting of distant to recent payoffs.[1]

[1]This mechanism for updating expected payoffs is a special case of criterion filtering, i.e., the direct updating of expected return functions on the basis of past return observations. See Tesfatsion (1979).

In an IPD/CR game, the players have some degree of control over the number of PD games they play—equivalently, over the number of moves they make—and players not participating in PD games can still receive wallflower and refusal payoffs. Different players can therefore end up playing different numbers of games, and not all payoffs are associated with game plays. Consequently, at the end of the tournament, we measure the *fitness* of each player by its average payoff score calculated as the total *sum* of its payoffs divided by the total *number* of its payoffs.

## 2.4 Finite State Machine Representations of Strategies

Following previous studies of the evolution of strategies for iterated prisoner's dilemma (Axelrod, 1987; Miller, 1989; Lindgren, 1991), each player in our simulations is uniquely associated with a fixed deterministic strategy for playing IPD against an arbitrary opponent an indefinite number of times, where the only information each player has about its current PD game partner is its play history with that partner.

We have implemented our model in two different programming languages (Turbo Pascal and C) using two different deterministic finite state machine structures: Moore machines as used in Miller (1989); and Mealy machines with a predetermined initial state, which we will refer to as IPD machines[2] and which are illustrated in Figure 1. Each of these structures allows a player to use a variable amount of its play history with each partner to determine its moves in plays with that partner.

The representation of IPD strategies affects various aspects of our evolutionary study, including the distribution of behaviors in the initial randomly-chosen population of players and the manner in which recombination and mutation create new players. Our use of two different representations has resulted in generally similar experimental outcomes, although some subtle differences have been detected. We have thus checked to some extent that our experimental outcomes are not simply artifacts arising from our particular choice of representation.

Unless otherwise noted, the simulation results reported in this paper are based on the IPD machine representation shown in Figure 1. However, many of these results have also been verified using Moore machines.

The genetic structure of each player is a bit string which encodes its IPD or Moore machine. In the floating $\tau$ - $\omega$ studies this genetic structure is augmented with either the $\tau$ or the $\omega$ values. This genetic structure determines its behavior under each possible set of circumstances, but players with different genetic structures can be observationally equivalent in the context of our IPD/CR game.

[2]Formally, the IPD machines are characterized by a six-tuple $(M, m_1, S, s_1, \lambda, f)$, where: $M \equiv \{c, d\}$ is the set consisting of the two possible moves for each PD game play, either cooperate $c$ or defect $d$; $m_1 \in M$ is an initial move; $S$ is a finite set of internal states; $s_1 \in S$ is the initial state resulting from the initial move $m_1$; $\lambda : S \times M \to M$ is an output function indicating the next move to be taken as a function of the current state and the current move of one's opponent; and $f : S \times M \times M \to S$ is a state transition function indicating the next state as a function of the current state, the current move of one's opponent, and the next move to be taken.

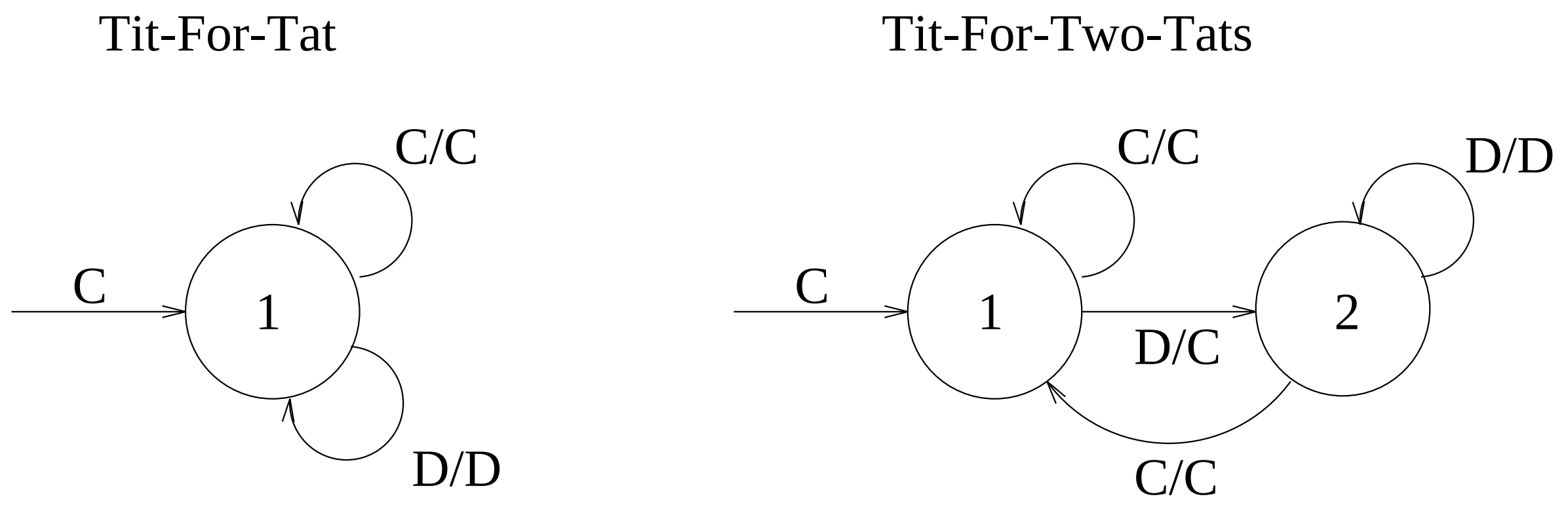

Figure 1: Two illustrative IPD machines, which are Mealy machines with fixed initial states. A player makes an initial move, either cooperate $c$ or defect $d$, and then enters the initial state 1; this initial move is indicated next to the arrow entering state 1. Once the player has arrived at a current state, its next move is conditioned on the previous move of its partner as well as on the current state. This move sequence then determines a transition to a new state. A transition to a new state is indicated by an arrow, and the move sequence or sequences that result in this transition are indicated beside the arrow in a move-slash-move format. The previous move of the partner appears to the left of the slashmark and the next move of the player appears to the right of the slashmark.

### 2.5 The Genetic Step

At the end of each tournament, the current generation of $N$ players is transformed into a new player population of the same size via a genetic algorithm that uses elitism, crossover, and mutation; and this new population then engages in another IPD/CR tournament.

More precisely, at the beginning of the genetic step, each player in the current population is assigned a fitness equal to its average payoff score per payoff received. Copies of the $X$ most fit players are retained in the next generation, and the bottom $N - X$ players are replaced by offspring of the top $X$ players.

Parents are selected by means of a roulette wheel selection (Goldberg, 1989). Two parents are selected $(N - X)/2$ times from the top $X$ players with a probability directly proportional to their relative fitness. A player is allowed to mate with itself.

The recombination (crossover) of two IPD machine parents is accomplished as follows. In the majority of our experiments, each machine has 16 states and is coded as a string consisting of 161 bits. (The first bit specifies the initial move, bits 2-129 are used for the 32 state transition arrows, each of which uses four bits to specify the next state, and bits 130-161 are used by the 32 arrow labels, which each use one bit to specify the move $c$ or $d$.) We generate a random variable $q$ that is distributed uniformly over the discrete range $1, 2, \ldots, 161$. The bits in positions $q$ through 161 of the parental bit strings are exchanged to obtain the bit strings for two offspring.

Next, the bit strings of these two offspring are subjected to mutation. For the initial move and the 32 arrow labels, each bit is flipped with probability $\mu$. For the 32 state transition arrows, *mutation of the state transition arrow* (i.e. all four bits) occurs with probability $\mu$; and, once a state transition arrow has been selected for mutation, a uniformly distributed random value is selected from the discrete range $1, 2, \ldots, 16$ and coded as a new four-bit representation for the state transition arrow. [3]

## 3 Distribution of Behavior in the Initial Population

Our player populations are small, implying that only a small subset of possible IPD strategies tends to be explored in each run. This is especially true given our small mutation rate. The genetic material available in the initial population is therefore one crucial factor determining the behavior of any specific run. In particular, at low values of $\omega$ the wallflower ecologies we describe in the next section are only observed when the expected number of mutually cooperative individuals in the initial population is small. Moreover, our initial set of IPD strategies is not chosen from a uniform distribution of all possible IPD strategies. Instead, selecting each arrow and each arrow label from a uniform distribution creates an initial distribution of IPD strategies that is highly biased towards simple strategies.

In this section we determine the probability that a randomly chosen IPD machine, as described in section 2.4, exhibits a given type of self play in an IPD game play against a clone of itself. We also determine the expected number of self-cooperators.

By construction, all moves in the self-play of an IPD machine are synchronized, either both $c$ or both $d$. Furthermore, some move and its associated state must eventually recur in self play, after which the machine will loop endlessly through the interim sequence of moves and states. The self-play behavior of these IPD machines can therefore be characterized by a *self-play string* of the form $A$:$B$, where $A$ and $B$ are strings consisting of $c$ and $d$ moves that are associated with state transitions. The string $A$ represents a series of moves (and state transitions) made initially in self play but not repeated, while $B$ represents a series of moves (and state transitions) that the IPD machine repeats thereafter. The string $A$ always contains at least the initial move since it is unrepeatable. Likewise, $B$ always contains at least one entry, a move whose associated state transition arrow points to the same state as the state transition arrow associated with the last entry in $A$.

To illustrate the point that each entry of a self-play string represents both a move and a state transition, consider the two self-play strings *c*:*c* and *c*:*cc*. Both represent pure self-cooperation, but an IPD machine with the former string uses only one state in its self play while one with the latter string uses at least two. This distinction is nontrivial at the level of evolution. An examination of their diagrams shows that the self-play string of an IPD machine with a *c*:*c* self-play string can only be affected by three point mutations (changing either the initial or the second $c$, or changing the destination of the arrow marked $c/c$ out of the initial state) while the *c*:*cc* self-play string can be modified by five different mutations.

**Lemma 1.** *There are $2 \cdot (2n)^{2n}$ distinct possible specifications for an $n$-state IPD machine.*

**Proof:**

The IPD machine has $n$ states and an initial move. For each state, the IPD machine must react to either a $c$ or a $d$ input, hence two arrows corresponding to these two possible inputs must be specified. An arrow must go to one of n states and be labeled with one of two moves, implying there are $2n$ different ways to specify a given arrow. The total number of arrows in an $n$-state IPD machine is $2n$, not counting the arrow pointing to the initial state, so we have $(2n)^{2n}$ ways to specify them. Multiplying this by the 2 possible initial moves gives the desired formula. □

[3] The mutation process was incorrectly reported in Stanley et al. (1994) as consisting solely of point mutations of the bit string.

**Theorem 1.**

Let $S{=}A{:}B$ denote a self-play string, and let $x, y, z \in \{c, d\}$ denote, respectively, the first character of $A$, the last character of $A$, and the last character of $B$. Let $n_c$ and $n_d$ denote the number of characters of type $c$ and $d$ in $S$, respectively. For any statement $b$, denote by $[b]$ the truth value of $b$: 1 if $b$ is true; 0 if b is false. Then the number of $n$-state IPD machines that have the self-play string $A{:}B$ is:

$$(2n)^{2n-n_c-n_d-[y\neq z]+1} \cdot \frac{n!(n-1)!}{(n-n_c+[z=c])!(n-n_d+[z=d])!} \ . \tag{3}$$

**Proof:**

We start by counting the number of ways to place the arrows labeled with the entries of the self-play string.

The first entry of the self play string involves no choice; its arrow points to state 1 and is labeled $x$. For each subsequent entry of the self-play string, apart from the last entry, the arrow labeled with that entry leads from the current state; and the state to which the arrow points may be freely chosen from those states that do not yet have an arrow with the same label pointing at them. The arrow for the last entry in the self-play string must point to the same state as the arrow corresponding to the move immediately preceding the colon in the self-play string.

This means we make an ordered choice of $(n_c-[x=c]-[z=c])$ states out of $(n-[x=c])$ to be the head states of the arrows corresponding to cooperation and an ordered choice of $(n_d-[x=d]-[z=d])$ states out of $(n-[x=d])$ to be the head states of the arrows corresponding to defection. The subtraction of $[x=c]$ and $[x=d]$ from the set of available states takes into account that the initial move uses up one of the moves that could otherwise be associated with the initial state. The subtraction of $[z=c]$ and $[z=d]$ from the set of arrows that need head states reflects the fact that the head state of the last arrow is dictated by the position of the colon in $S$. Since states may be the heads of a $c$ arrow and/or a $d$ arrow, the choices are independent and the total number of choices is

$$\frac{(n-[x=c])!(n-[x=d])!}{(n-n_c+[z=c])!(n-n_d+[z=d])!} \ . \tag{4}$$

Since exactly one of $[x=c]$ and $[x=d]$ is 1, the numerator can be simplified to $n!(n-1)!$. This transforms the above to

$$\frac{n!(n-1)!}{(n-n_c+[z=c])!(n-n_d+[z=d])!} \ . \tag{5}$$

Any $n$-state IPD machine has a total of $2n$ arrows, not counting the arrow associated with the initial move. We explicitly chose the description of $n_c+n_d-1$ of these arrows in the course of laying out the self-play string. If $y=z$, then we have chosen enough arrows to ensure that the IPD machine will exhibit the desired self play; but, if $y \neq z$, then we must ensure that the head state of the state transition arrow associated with $y$ transits to the same next state, and is labeled with the same move, regardless of whether the previous move was a $c$ or a $d$ (i.e., regardless of whether this head state is being entered from $y$ or from $z$). This requires we fill in one additional arrow description.

Thus, the total number of arrow descriptions filled in during the course of specifying the self-play string is $n_c + n_d + [y \neq z] - 1$. This leaves $2n - n_c - n_d - [y \neq z] + 1$ arrow descriptions to be filled in. Since there are $2n$ different ways to fill in each arrow description, we see that there are

$$(2n)^{2n-n_c-n_d-[y\neq z]+1} \tag{6}$$

additional choices not associated with specifying the self-play string. Multiplying the choices yields the desired formula (3). □

**Corollary 1.**

If $n_c$, $n_d$, $x$, $y$, and $z$ are as in Theorem 1, then the probability of a randomly-generated IPD machine having a given self-play string $A{:}B$ is

$$\frac{n!(n-1)!}{2 \cdot (2n)^{(n_c+n_d+[y\neq z]-1)}(n-n_c+[z=c])!(n-n_d+[z=d])!} \; . \tag{7}$$

**Proof:**

Divide equation 3 by the total number of possible IPD machines computed in Lemma 1. □

Call an IPD machine *self-cooperative* or *self-defecting* if its self-play string consists entirely of $c$ or $d$ moves, respectively. The next result is valid for both self-cooperative and self-defecting IPD machines:

**Theorem 2.**

The probability that a randomly-generated IPD machine is self-cooperative is

$$\sum_{l=1}^{n} \frac{(n-1)! \cdot l}{2 \cdot (2n)^l (n-l)!} \; . \tag{8}$$

**Proof:**

With an initial move that must come before the colon, and up to $n$ states corresponding to other potential moves in the self-play string, at least one of which must come after the colon, we see that the number $l$ of entries in a self-play string consisting entirely of $c$ moves is in the range $2 \leq l \leq n+1$. Moreover, in a string with $l$ such entries, there are $l-1$ places to put the colon.

The index of summation will run across the number of possible entries in a completely cooperative self-play string, minus one. For a given $l$ and colon position, formula (7) may be filled in with $n_c = l$, $n_d = 0$, and $x = y = z = c$ to yield the probability

$$\frac{n!(n-1)!}{2 \cdot (2n)^{(l-1)}(n-l+1)!(n)!} \; . \tag{9}$$

Since there are $l-1$ places the colon could be placed, the probability of an IPD machine having a self-play string with $l$ cooperates and no defects is, after cancellation,

$$\frac{(n-1)!(l-1)}{2 \cdot (2n)^{(l-1)}(n-l+1)!} \; . \tag{10}$$

| $n$ | *Probability* | *Expectation* | $n$ | *Probability* | *Expectation* |
|---|---|---|---|---|---|
| 1 | 0.25 | 7.5 | 12 | 0.0615 | 1.8456 |
| 2 | 0.1875 | 5.625 | 13 | 0.0578 | 1.735 |
| 3 | 0.1528 | 4.5833 | 14 | 0.0546 | 1.6373 |
| 4 | 0.1299 | 3.8965 | 15 | 0.0517 | 1.5502 |
| 5 | 0.1134 | 3.402 | 16 | 0.0491 | 1.4721 |
| 6 | 0.1009 | 3.0259 | 17 | 0.0467 | 1.4016 |
| 7 | 0.091 | 2.7287 | 18 | 0.0446 | 1.3377 |
| 8 | 0.0829 | 2.4872 | 19 | 0.0426 | 1.2795 |
| 9 | 0.0762 | 2.2866 | 20 | 0.0409 | 1.2262 |
| 10 | 0.0706 | 2.1171 | 25 | 0.0338 | 1.0154 |

Table 2: The probability that a randomly generated $n$-state IPD machine will be self-cooperative, and the expected number of self-cooperative ones out of thirty.

If we then sum over the possible values of $l$ and correct the index to run from 1 to n, we obtain formula (8). □

For later purposes, Table 2 computes the expected number of self-cooperative $n$-state IPD machines in a randomly-chosen population of size thirty.

# 4 Simulation Findings: Overview

In preliminary IPD/CR simulation studies reported in Stanley et al. (1994), we found that the overall emergence of cooperation was faster with choice and refusal than with round-robin partner selection as used in Miller (1989). As well, the initial dip in overall average fitness was either much shallower or missing entirely. Although a large number of ecologies emerged that were essentially mutually cooperative, other ecologies had an average fitness that remained nearly constant for many generations at levels below the mutual cooperation payoff, or which displayed other fairly regular payoff patterns. Interestingly, when we overlaid the average fitnesses from different runs at the same parameter values, the average fitnesses often clustered around a small number of values.

These general findings continue to hold for the more extensive studies reported in the present paper. Indeed, the enhancement of the emergence of cooperation appears even more dramatic when we compare our results to random rather than to round-robin partner selection, a more appropriate comparison because the total number of games played is more similar. In this section we present a brief overview of these and other simulation findings. A more detailed discussion is given in Section 5.

When partners are chosen randomly and no refusal is allowed, many populations never evolve to full cooperation. On the other hand, only a small fraction evolve to full defection; the rest go to meta-stable states which lie between these two cases. As depicted in Figure 2(b), these effects show up visually as either noisy horizontal lines or thin bands when the average fitnesses of each of the forty runs are graphically superimposed. Some of these are easy to understand: 3.0 reflects fully cooperative play; 2.8 is a population in which each player defects exactly once against each other player; and so on. The overall average fitness (i.e., the average of the average fitness across the forty runs) stays well below 3.0, the full-cooperation payoff (see Figure 2(a)).

| | |
|---|---|
| Number of Players | $N = 30$ |
| Number of Generations (Tournaments) | $G \geq 50$ |
| Number of Iterations per Tournament | $I = 150$ |
| Initial Expected Payoff: | $\pi_0 = 3.0$ |
| Minimum Tolerance Level: | $\tau = 1.6$ |
| Refusal Payoff: | $R = 1.0$ |
| Wallflower Payoff: | $W = 1.6$ |
| Memory Weight: | $\omega = 0.7$ |
| Number of Elite | $X = 20$ |
| Mutation Probability | $\mu = 5/1000$ |

Table 3: Parameter Settings for the Standard IPD/CR Scenario

Contrast this with Figure 3, which depicts the average fitnesses of forty different runs at the standard IPD/CR parameter settings specified in Table 3. Once again, not all populations evolve to full cooperation. However, nearly all of these partially noncooperative populations evolve to a situation where, on average, each player defects only once against each other player. Thus, while horizontal bands are evident in both Figure 2(b) and Figure 3(b), the average fitnesses for the latter case tend to cluster in just two narrow regions; and the overall average fitnesses achieved for the standard IPD/CR scenario are higher than for random choice.

Figure 3 illustrates a commonly observed configuration for average fitnesses in the evolutionary IPD/CR: values clustered fairly tightly into a small number of narrow regions. For some parameter settings, however, the number of of these regions can be large and more diffusely distributed. Also, choice and refusal can lead to situations where the players defect against each other a small number of times and then cease game play altogether, becoming wallflowers. The latter situation happens, for example, when the standard IPD/CR scenario is perturbed in any of the following three ways: the refusal payoff is increased until being refused becomes an attractive alternative to mutual defection; the memory weight is lowered until cooperators refuse to tolerate even a single defection; or the wallflower payoff is increased to a level lying only slightly below the mutual cooperation level. The first two cases are illustrated by Figures 4 and 5, respectively.

Despite attaining similar average fitnesses, the populations with average fitnesses in any one region consist of genetically diverse players with interaction patterns peculiarly adapted to the choice and refusal mechanism. For example, as described further in Section 5, the source of the spiking in average fitnesses from 2.69 to 3.0 observed in Figure 3 is an intricate dance between one set of initially rapacious but ultimately love-struck players ("Bobs") and another set of temptingly cooperative players ("Raquels"). A more detailed examination of the fascinating social network structures arising in the evolutionary IPD/CR is given in a companion paper (Smucker et al., 1994). These structures are strongly reminiscent of the social network structures observed in real world settings.

If the choice and refusal parameters $\tau$ and $\omega$ are allowed to evolve over time along with the players' IPD strategies, then populations typically evolve to be intolerant of anyone who defects even once against a cooperator. Interestingly, as discussed more carefully in Section 5, the populations with evolved $\tau$ and $\omega$ values tend to achieve average fitnesses lying between those of populations with fixed choice and refusal parameters and those attained by populations using random partner choice. However, the variety of behaviors observed in the populations with evolved $\tau$ and $\omega$ values is just as great as for the populations with fixed choice and refusal parameters.

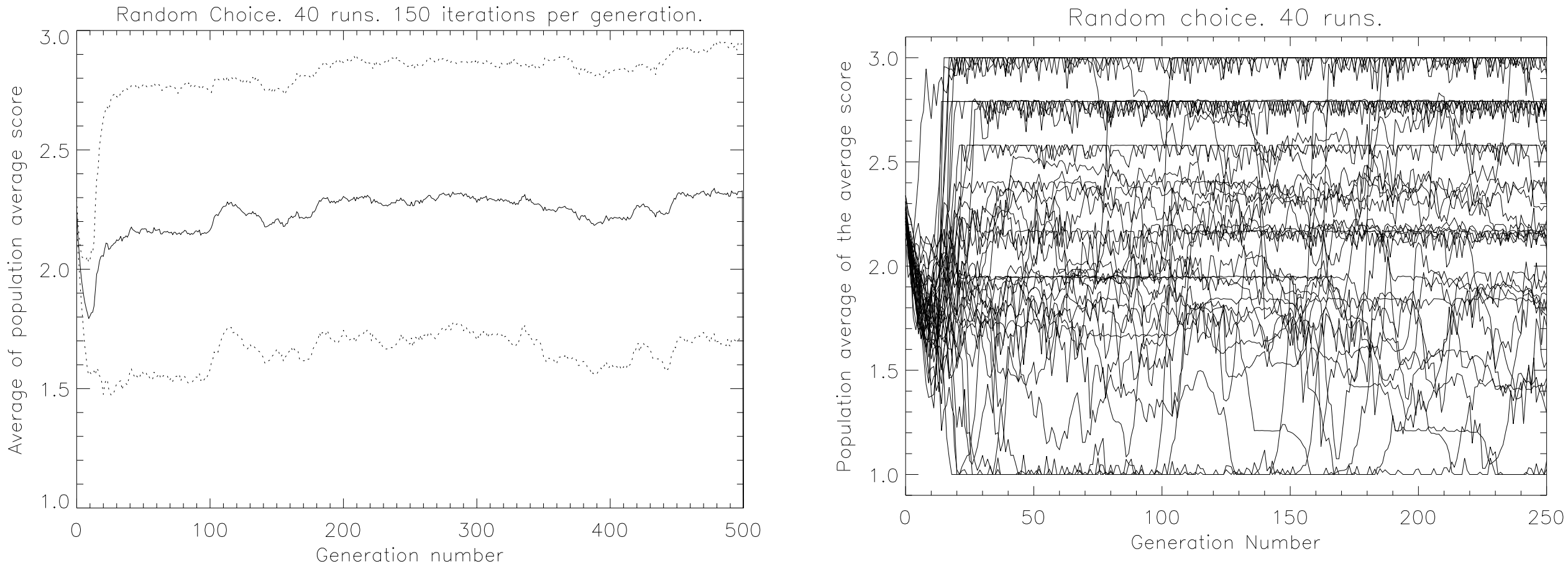

Figure 2: Random choice evolved for 500 generations. Each player chooses exactly one partner at random on each of the 150 iterations comprising an IPD tournament. (a) The overall average fitness achieved by successive generations across 40 runs. The dashed lines (error bounds) show this overall average fitness plus or minus one standard deviation. (b) Each line shows the average fitness achieved by successive generations during one of the 40 runs. Note the wide spread and the horizontal bands. The bands tend to occur because populations become genetically homogeneous and mutants tend to do poorly.

# 5 Simulation Studies: Detailed Results

In this section we detail a variety of simulation studies that have been conducted to explore the sensitivity of IPD/CR evolutionary outcomes to changes in key parameters. We start by explaining the concept of a fitness region, used below to aid the reporting of our simulation findings.

## 5.1 Fitness Regions

As seen in Section 4, the average fitnesses achieved by successive generations in runs of the evolutionary IPD/CR from different initial random seeds tend to cluster around a small number of levels. We refer to these levels as *regions* $D^0$, $D^1$,... . Region $D^n$ is the cluster of average fitnesses roughly centered around the average fitness achieved by a homogeneous (genetically identical) population whose self-play string $A{:}B$ consists of an initial string $A$ of $n$ defections followed by a string $B$ consisting entirely of cooperations. Thus, region $D^0$ is the fitness cluster roughly centered around the average fitness 3.0 of a homogeneous population with a self-play string $c{:}c \dots c$, region $D^1$ is roughly centered around the average fitness of a homogeneous population with self-play string $d{:}c \dots c$, and so forth.

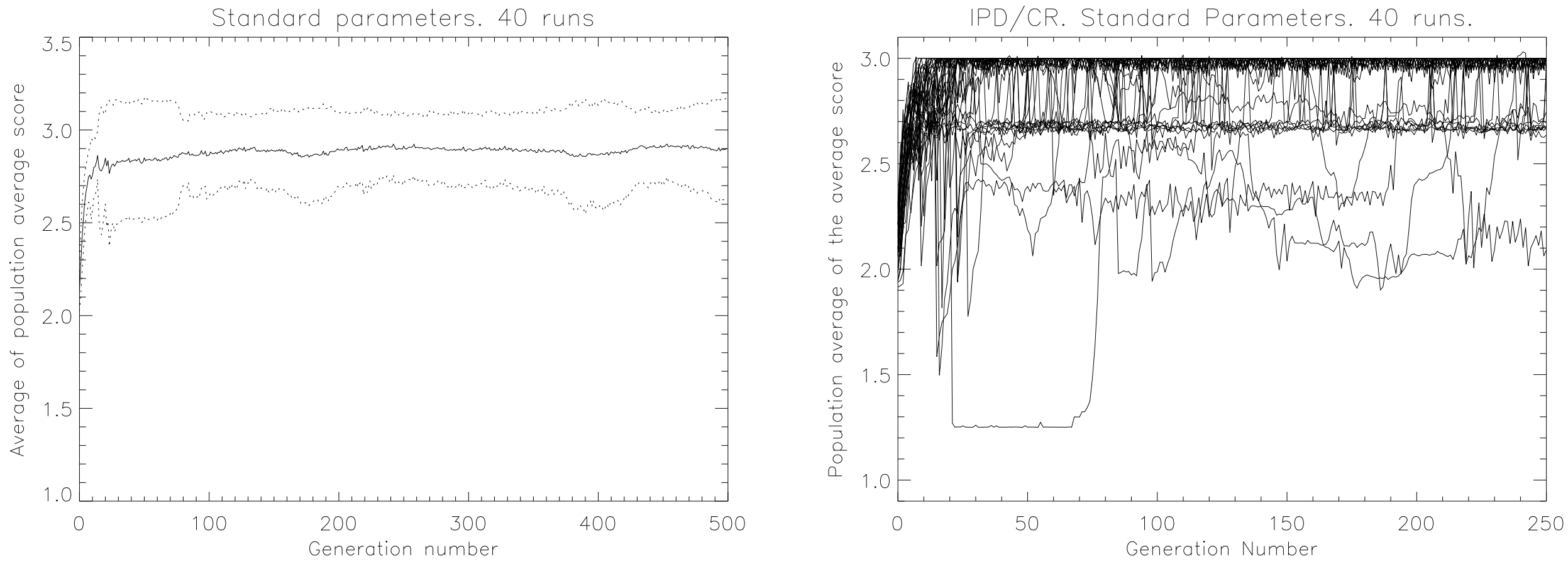

Figure 3: IPD/CR evolved for 500 generations with all parameters at their standard scenario levels. As in Figure 2, each player chooses at most one partner in each iteration. (a) Overall average fitness across forty runs and error bounds. (b) Average fitness achieved by successive generations for 40 individual runs. Note how few fitness levels are achieved in comparison to Figure 2. The jumps in average fitness from the fitness region near 2.69 to a level above the mutual cooperation fitness region at 3.0 are observed frequently, and indicate the Raquel-and-the-Bobs phenomenon discussed in the text.

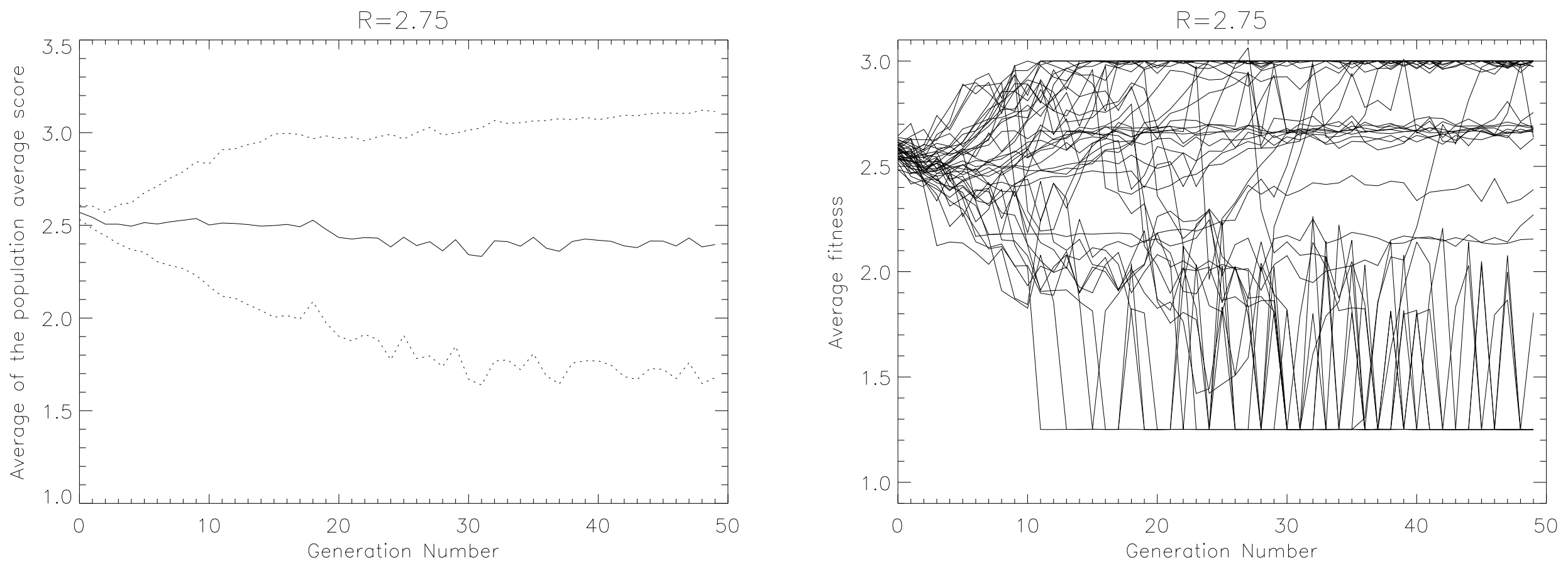

Figure 4: IPD/CR with a high refusal payoff ($R = 2.75$.). All other parameter settings are at standard scenario levels. (a) Overall average fitness across forty runs and error bounds. (b) Average fitnesses for forty individual runs. The high refusal payoff is so attractive that some runs gets trapped in a wallflower situation where each player defects against each other player four times and thereafter collects wallflower payoffs. Every time a more cooperative player appears, the defectors take advantage of it to collect refusal payoffs; the cooperator itself scores poorly and immediately dies out. The appearance of such cooperators is indicated by spikes in the average fitnesses.

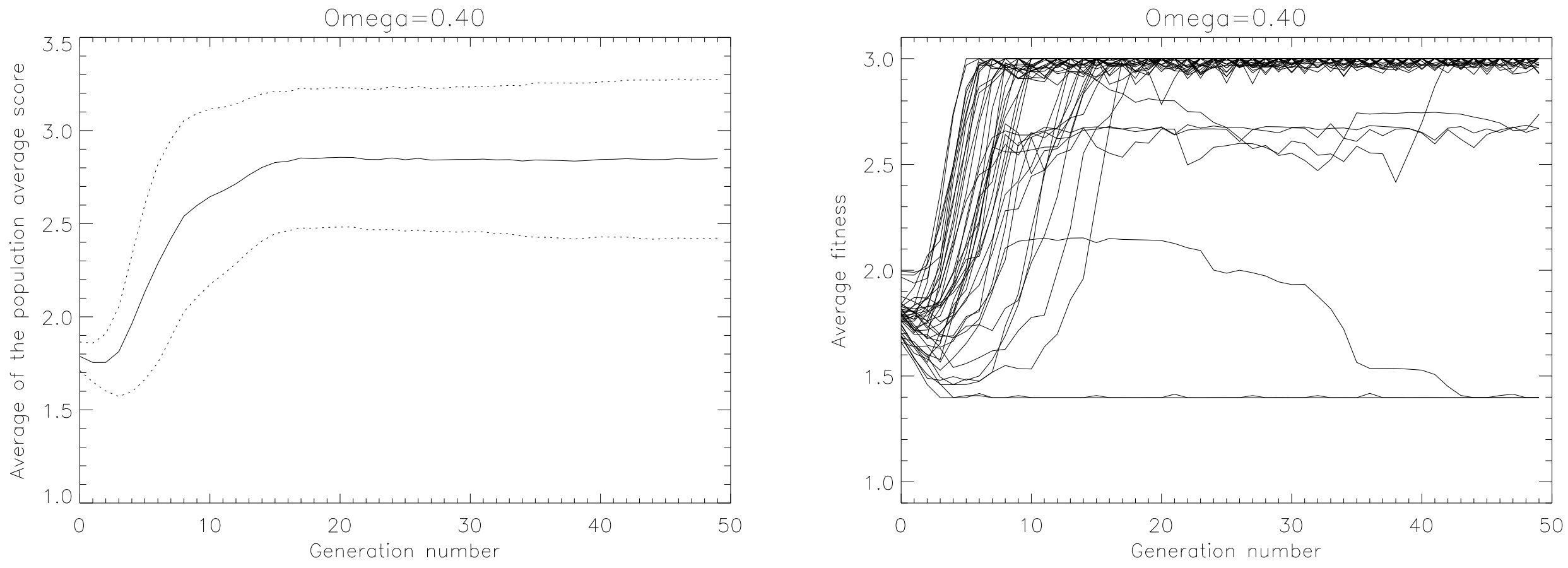

Figure 5: IPD/CR with low memory weight ($\omega = 0.4$). All other parameter settings are at standard scenario levels. (a) Overall average fitnesses across forty runs and error bounds. (b) Average fitnesses for forty individual runs. Here cooperators do not accept any defections, and mutual defectors only play each other twice before preferring to become solitary. Also, some populations become trapped in wallflower situations for many generations, with mutant cooperators achieving even lower payoffs than their defecting relatives.

Note that the precise average fitness anchoring a region $D^j$ with $j > 0$ depends on the IPD/CR parameter settings. For example, given the standard settings in Table 3 with $I$ = 150 iterations between genetic steps, the average fitness anchoring $D^1$ is 2.69; but this average fitness increases monotonically to 3.0 as $I$ becomes arbitrarily large. Also, the player populations of the ecologies lying within a particular fitness region can vary enormously from one another; and, even for any one of these ecologies, the successive generations are rarely homogeneous. For example, region $D^1$ may contain ecologies consisting of a homogeneous population of players having a self-play string *cd*:*c*, ecologies consisting of a mixed population of players having self-play strings *d*:*c* and *dc*:*c*, as well as ecologies consisting of a homogeneous population having a self-play string *d*:*c*

Finally, the fitness cluster centered roughly around the wallflower payoff $W$ will be referred to as the *wallflower region.* The ecologies falling in this region typically consist of player populations that initially engage in mutual defections and ultimately end up as wallflowers.

## 5.2 Sensitivity to Changes in the IPD/CR Parameters

A series of sensitivity experiments were conducted for various subsets of the IPD/CR parameters, keeping all remaining parameters at their standard IPD/CR scenario settings as listed in Table 3. In all experiments except the floating $\tau$ case reported later, the wallflower payoff $W$ was set equal to $\tau$.[4] As will be seen below, an important implication of these sensitivity experiments is that the IPD/CR parameters $\tau$, $\omega$, and $\pi_0$ have closely coupled effects on the evolution of player populations.

[4] Intuitively, it makes little sense for a player to refuse (accept) an offer whose expected payoff is higher (lower) than the expected payoff $W$ from cessation of game play.

Figure 6 describes how the mean and standard deviation of average fitnesses change when the minimum tolerance level $\tau$, the memory weight $\omega$, the initial expected payoff $\pi_0$, and the refusal payoff $R$ are varied one at a time from their standard scenario settings. For each tested parameter configuration, forty runs were made using forty different initial random seeds, resulting in forty distinct ecologies, and each run consisted of fifty generations of players. The mean average fitnesses $\bar{m}$ for the final twenty-five generations are indicated by squares, and the dispersion of each mean $\bar{m}$ is indicated by an error bar giving a range of plus or minus one standard deviation $\bar{\sigma}$ about this mean.[5]

As seen in part (d) of Figure 6, $\bar{m}$ drops precipitously in response to increases in the refusal payoff, $R$, because nearly all populations evolve into wallflower ecologies. Surprisingly, however, parts (a) through (c) indicate that increases in $\tau$, $\omega$, or $\pi_0$ have little effect on $\bar{m}$. In particular, the player populations for the final twenty-five generations are more cooperative on average than the populations evolved with random partner choice; compare Figure 2. As seen in part (b), this remains true even when the memory weight $\omega$ is set at 0.9, implying that players only gradually move away from the common expected payoff $\pi_0$ they initially have for all potential game partners.[6]

In contrast, the average standard deviation, $\bar{\sigma}$, tends to increase with increases in $\pi_0$ or $R$, and also to vary significantly in response to increases in $\tau$ and $\omega$, first decreasing and then increasing. These findings are consistent with the following observations. With low $\tau$, high $\omega$, and high $\pi_0$, although $D^0$ and $D^1$ fitness regions are evident, many ecologies persist outside these regions in no discernible pattern. As illustrated in Figure 7, this results in a large dispersion in average fitnesses. In contrast, for more intermediate parameter settings, near the standard scenario values and for low $\pi_0$, almost all ecologies lie within either the $D^0$ or the $D^1$ fitness regions, implying a relatively smaller dispersion in average fitnesses. Finally, for high $\tau$ and low $\omega$, the ecologies tend to divide between the cooperative fitness region $D^0$ and the wallflower fitness region, resulting once again in a rather large dispersion in average fitnesses.

Table 4 provides a more detailed description of the one-parameter sensitivity outcomes summarized in Figure 6 for $\tau$, $\omega$, $\pi_0$, and $R$. Caution must be exercised in interpreting these results; the somewhat different outcomes observed in the four different forty-run experiments undertaken at standard scenario parameter settings (see $\tau = 1.6$, $\omega = 0.7$, $\pi_0 = 3.0$, and $R = 1.0$) indicate that our sample size is too small to ensure that all interesting phenomena that can occur at a particular parameter setting are actually in evidence. Table 5 reports results for a two-parameter sensitivity study in which $\tau$ and $\pi_0$ were varied together. As before, forty runs were made for each tested parameter configuration, and each run consisted of fifty generations.

[5]The means and standard deviations $\bar{m}$ and $\bar{\sigma}$ are determined in the following manner. For each parameter setting, we first determine the average fitness $m(e, g)$ attained by each of the forty ecologies $e$ during each generation $g = 1, \ldots, 50$. We then calculate the mean $m(g)$ of the average fitnesses $m(e, g)$ across the forty ecologies $e$ for each $g$. As seen in Figures 2(a)-5(a), the means $m(g)$ tend to level out by about the twenty-fifth generation, with only small subsequent variations. Consequently, we determine the overall mean $\bar{m}$ by taking the average of the means $m(g)$ across generations $g = 25, \ldots, 50$. Also, for each parameter setting, we calculate the average standard deviation $\bar{\sigma}$ of the average fitnesses $m(e, g)$ achieved by the forty ecologies $e$ over generations $g = 25, \ldots, 50$.

[6]When $\omega$ is set at 1.0, all players remain indifferent concerning their choice of partner since the initial expected payoff, $\pi_0$, is never updated. The value $\omega = 1.0$ was used to generate the random choice results reported in Figure 2.

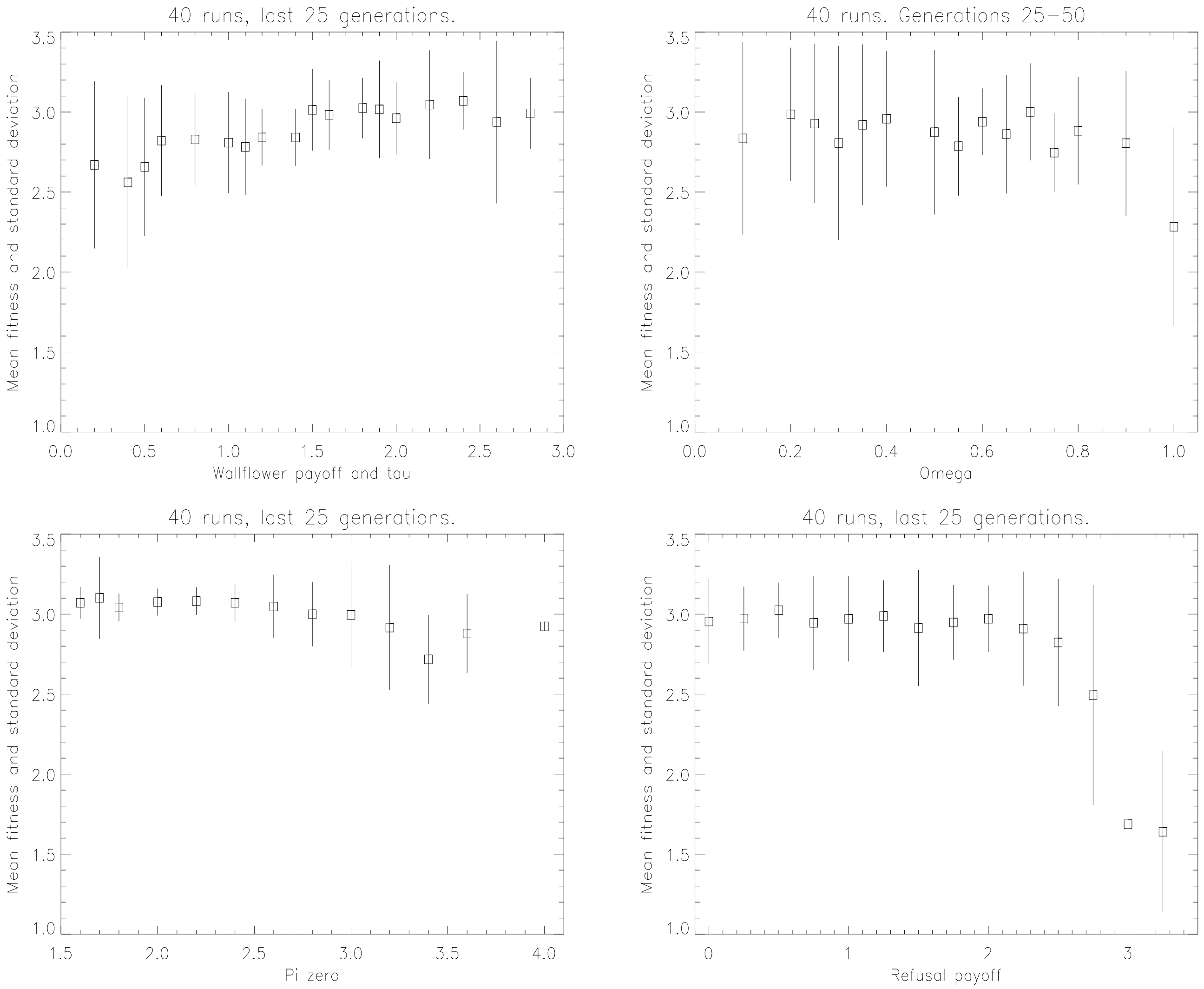

Figure 6: Sensitivity of average fitnesses to parameter variations. For each indicated parameter setting, the square indicates the mean average fitness achieved by the forty ecologies over the final twenty-five generations, and the error bar gives a (two standard deviations) range of dispersion about this mean.

| $\tau = W$ | minimum | phenomena | $\omega$ | minimum | phenomena |
|---|---|---|---|---|---|
| 0.2 | 1.0 | $N, D^0, nD^1, W$ | 0.10 | 1.5 | $D^0, W$ |
| 0.4 | 1.0 | $N, D^0, D^1, D^2, L_2$ | 0.20 | 1.5 | $D^0, W$ |
| 0.6 | 1.0 | $D^0, nD^1, N, O_1$ | 0.25 | 1.5 | $D^0, W$ |
| 0.8 | 1.3 | $D^0, nD^1, N, D_1^2$ | 0.30 | 1.4 | $D^0, D^1, W$ |
| 1.0 | 1.5 | $D^0, nD^1, N$ | 0.35 | 1.4 | $D^0, D^1, W$ |
| 1.2 | 1.6 | $N, D^0, nD^1, C_1$ | 0.40 | 1.4 | $D^0, nD^1, W$ |
| 1.4 | 1.3 | $N, D^0, nD^1, J_2$ | 0.50 | 1.4 | $D^0, D_1^1, O_2, W, C_1$ |
| 1.6 | 1.7 | $D^0, nD^1, J_2$ | 0.60 | 1.3 | $D^0, D^1, D_2^2, D_1^3, L$ |
| 1.8 | 1.6 | $D^0, nD^1, J, D_1^2$ | 0.65 | 1.4 | $N, nD^1, C_1$ |
| 2.0 | 1.7 | $nD^0, nD^1, J, O, C_1$ | 0.70 | 1.6 | $D^0, nD^1, J_1$ |
| 2.2 | 1.8 | $D^0, D_1^1, W_1$ | 0.75 | 1.3 | $N, nD^0, nD^1, nD^2$ |
| 2.4 | 1.9 | $D^0, W_1$ | 0.80 | 1.6 | $nD^1, nD^0, nD^2$ |
| 2.8 | 2.4 | $D^0, W$ | 0.90 | 1.3 | $N, nD^1, nD^0$ |
| $\pi^0$ | | | $R$ | | |
| 1.6 | 2.5 | $D^0, D^1$ | 0.00 | 1.2 | $D^0, D^1, N, J$ |
| 1.8 | 2.4 | $D^0, W_1, O_1$ | 0.25 | 1.5 | $D^0, nD^1, J, D_1^4$ |
| 2.0 | 2.0 | $D^0$ | 0.50 | 1.3 | $D^0, nD^1, J$ |
| 2.2 | 2.2 | $D^0, D_2^1$ | 0.75 | 1.3 | $D^0, nD^1, J, N$ |
| 2.4 | 2.3 | $nD^0$ | 1.00 | 1.6 | $D^0, nD^1, D_1^2, J_4, O$ |
| 2.6 | 1.4 | $nD^0, D_1^1$ | 1.25 | 1.3 | $D^0, D^1, D_1^2, J_2, O_2$ |
| 2.8 | 1.6 | $D^0, D_1^1, N$ | 1.50 | 1.3 | $D^0, D^1, J, N$ |
| 3.0 | 1.6 | $D^0, nD^1, J_2, O_1$ | 1.75 | 1.4 | $D^0, nD^1, D_1^2, N, J_2$ |
| 3.2 | 1.2 | $D^0, D^1, L_1, O_1$ | 2.00 | 1.5 | $D^0, nD^1, N$ |
| 3.4 | 1.2 | $D^0, D^1, D_2^2, N$ | 2.25 | 1.2 | $N, D^0, nD^1, W_1$ |
| 3.6 | 1.2 | $D^0, nD^1, nD^2$ | 2.50 | 1.2 | $D^0, nD^1, N, W_2$ |
| 4.0 | 1.6 | $D^0, nD^1, N$ | 2.75 | 1.2 | $W, D^0, D^1, D_2^3, D_1^2, J$ |
| | | | 3.00 | 1.2 | $W, D_1^0, N$ |

Table 4: One-parameter sensitivity results. "Minimum" is a rough estimate of the minimum average fitness achieved by the forty ecologies after an initial transient stage, and gives a crude sense of the level of cooperation evinced. The phenomena columns indicate the various behaviors observed for the forty ecologies, roughly in order of their frequency. $D^n$ = fitness region $n$, $W$ = wallflower region, $C$ an ecology which persists for a long time whose dominant subpopulation has a self-play string with a short cyclic section, and $O$ = other ecologies with nearly constant fitness across many generations. Also, $N$ = noisy average fitness plots with no discernible patterns, $J$ = spiking observed in average fitness, and $L$ = a late-appearing wallflower ecology. A small $n$ in front of a phenomenon indicates that it was evident amidst some amount of noise, and a subscript $m$ on a phenomenon indicates that only $m$ cases were observed.

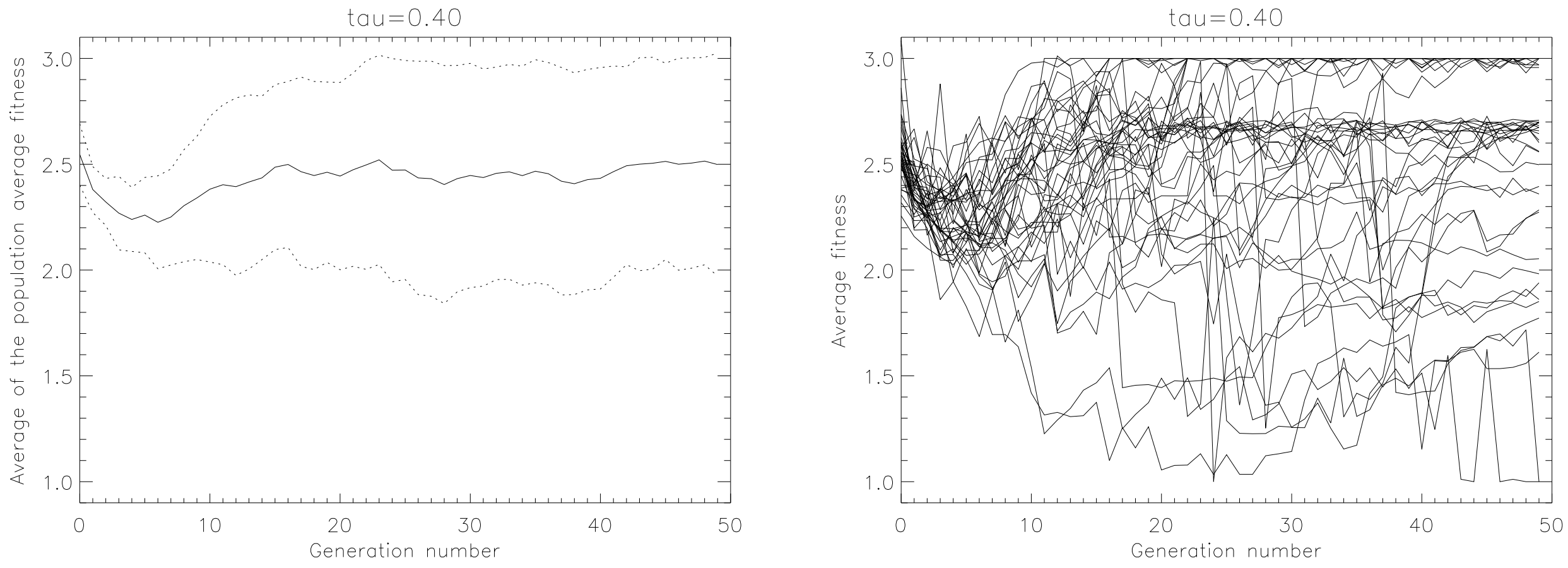

Figure 7: Experimental findings with low minimum tolerance level ($\tau = 0.40$). All other parameter settings are at standard scenario levels. (a) Overall average fitness across forty runs, and error bounds. (b) Average fitnesses for forty individual runs.

| $\pi_0 \backslash \tau$ | 1.6 | 2.0 | 2.4 | 2.8 |
|---|---|---|---|---|
| 1.6 | 2.6<br>$D^0$ | | | |
| 2.0 | 1.9<br>$nD^0$ | 1.8<br>$D^0, W$ | | |
| 2.4 | 1.9<br>$nD^0$ | 1.7<br>$D^0, W$ | 2.1<br>$D^0, W$ | |
| 2.8 | 1.7<br>$nD^0, N$ | 2.3<br>$D^0, nD^1$ | 2.1<br>$D^0, W, O_2$ | 2.5<br>$D^0, W$ |
| 3.0 | 1.7<br>$D^0, D^1, C_1, J$ | 1.6<br>$D^0, D^1, L, J$ | 1.9<br>$D^0, W, D^1_1$ | 2.5<br>$D^0, W$ |
| 3.2 | 1.3<br>$D^0, nD^1, N$ | 1.6<br>$D^0, nD^1$ | 1.9<br>$D^0, W, D^1_2$ | 2.5<br>$D^0, W$ |

Table 5: Two-parameter sensitivity results: $\tau$ and $\pi_0$ varied together. See the caption of Table 4 for a key. Each box reports the minimum average fitness for generations 25-50 as well as key observed phenomena.

The parameter $\pi_0$ determines the initial reactions of the players to each other. As seen in Tables 4 and 5, significant changes in evolutionary outcomes often occur when $\pi_0$ deviates from the mutual cooperation payoff 3.0. A value of $\pi_0$ greater than 3.0 encourages players to experiment by playing games with many new partners, while a value of $\pi_0$ lower than 3.0 encourages players to stick with those they have already played.

The exogenously chosen initial expected payoff $\pi_0$ can be dead wrong as an assessment of a potential partner. Therefore, a player's memory as embodied in its current expected payoffs can be tantamount to fantasy. The memory weight $\omega$, which weights past expected payoffs relative to newly obtained payoffs in each player's updating algorithm (2), can thus play an important role in offsetting or amplifying unwarranted optimism or pessimism stemming from an inappropriate setting for $\pi_0$.

In particular, given standard scenario settings for other parameter values, dramatic changes occur in evolutionary outcomes when $\omega$ is set low enough that any defection against a cooperation results in immediate refusal of all future PD-game offers. As seen in Table 4, the emergence of wallflower ecologies is then common. On the other hand, $\omega = 1$ results in random partner choice, and it might therefore be anticipated that the behaviors observed at high $\omega$ values will mimic those observed for random partner choice. However, it is only at the very highest tested value of $\omega$, 0.9, that we see any hint of the wide dispersion in average fitnesses that occurs for random partner choice, as depicted in Figure 2(b); and, even for this high $\omega$ value, no ecologies inhabiting the fitness regions $D^n$ with $n > 2$ are observed.

The sensitivity results reported in Table 4 for the refusal payoff $R$ reveal two regions of distinct behavior, splitting roughly at $R = 2.0$. When $R$ is below 2.0, refusal is avoided by the players and wallflower ecologies are absent. As $R$ increases, however, players tantalized by high $R$ payoffs often defect their way into a wallflower ecology. Thereafter, they only obtain $R$ payoffs when the population is invaded by a mutant cooperator.

In the next several subsections we examine more closely the behavioral phenomena highlighted in Tables 4 and 5.

### 5.2.1 The Cooperative Fitness Region $\mathbf{D}^0$

Ecologies whose average fitnesses lie in the $D^0$ region consist largely of players engaging in mutual cooperation, apart from an occasional mutant. As indicated in Tables 4 and 5, such ecologies have appeared for almost every parameter setting we have tested.[7] Indeed, for most parameter settings, a substantial number of ecologies approach this fitness region within the first fifteen generations. In general, the greater the fraction of ecologies that lie within the fitness region $D^0$, the more time they spend there and the higher is their attained average fitness, $\bar{m}$.

When all players engage in mutual cooperation, only three types of player interaction patterns are possible. If the initial expected payoff $\pi_0$ equals the mutual cooperation payoff 3.0, then potential PD game partners always have an expected payoff equal to 3.0 and each player is indifferent concerning whom it plays. In this case the partner selection mechanism reduces to random choice. If $\pi_0$ is greater than 3.0, players keep selecting new partners in a round-robin fashion as they experience disappointment from their lower than expected mutual cooperation payoffs. If $\pi_0$ is less than 3.0, each player latches onto the first individual it plays since its updated expected payoff for this partner will rise above the payoff $\pi_0$ it expects from each other partner.

[7] For the one-parameter sensitivity results, the only exceptions were when $\omega = 0.65$ or when $R$ was set higher than 3.0; for $R = 3.0$, only one such ecology appeared in forty runs. No exceptions were found for the two-parameter sensitivity studies for $\pi_0$ and $\tau$.

The $D^0$ fitness region is visually noisy when ecologies persistently move in and out of the region, or when other ecologies persist at nearby average fitness levels. This situation is indicated by $nD^0$ in Tables 4 and 5.

### 5.2.2 The Wallflower Fitness Region

The next most easily understood fitness region is the wallflower region. The ecologies falling within this region typically comprise players that initially defect against all other players, ultimately decide that each other player is intolerable, and thereafter collect only wallflower payoffs. Such ecologies are easily detected because they persist for a long time with an average fitness that is near the wallflower payoff $W$.

Two situations encourage wallflower ecologies to emerge and persist: positive incentive (a high $W$ or $R$ value); and quick refusal of defectors (a high $\tau$ value). In simulation experiments in which neither of these holds, we rarely observe the emergence of wallflower ecologies. On the other hand, when $R$ is set sufficiently high, almost all ecologies in our sensitivity experiments evolve into wallflower ecologies. Also, keeping $W$ set equal to $\tau$, and setting $\tau$ high, some (but not all) ecologies evolve into wallflower ecologies. It may seem counterintuitive that a high $R$ value results in more wallflower ecologies than a high $\tau$ value since synchronized play behavior is commonly observed, and players engaging in such behavior never receive refusal payoffs from each other. However, with a high $R$ value, defectors score very well whenever cooperators are present. Thus defectors tend to take over the population and prevent mutant cooperative players from invading.

Given $\pi_0 = 3.0$ and $\tau > 2.1$, with $\omega$ set at its standard scenario value of 0.7, a cooperating player will immediately refuse all further play with a player who defects on its first move against it. This immediate refusal in response to an initial sucker payoff of 0 increases the probability that a wallflower ecology will emerge. It also suggests why, in general, wallflower ecologies are observed in the two-parameter experiments of Table 5 only for the higher values of $\tau$ for each given value of $\pi_0$; for the appearance of wallflower ecologies in this table is roughly tracing out the boundary in the $\pi_0$–$\tau$ plane between immediate refusal and no immediate refusal in response to an initial 0 payoff, given $\omega = 0.7$.

When $\pi_0 < 3.0$, however, the explanation for the emergence of wallflower ecologies is actually more subtle than this discussion suggests, for a player's expected payoff then increases with each new mutual cooperation payoff, 3, that it receives. Thus, a sucker payoff, 0, or a mutual defection payoff, 1, received on the *first* move with another player might result in refusal of all further play, but refusal of further play might not occur if such payoffs are only received following a string of mutual cooperation payoffs. For example, as indicated in Table 5, wallflower ecologies were not observed when $\pi_0$ and $\tau$ were both set at 1.6, even though a 0 payoff on either the first or the second move always evokes immediate refusal in this case (but a 0 payoff on the third move need not).

In general, our findings suggest that wallflower ecologies primarily occur in the region of the parameter space for $\tau$, $\omega$, and $\pi_0$ where players are relatively intolerant of defections. In order to quantify this intolerance region, let $\pi(Z)$ denote the expected payoff that a player $i$ has for another player $j$ after receiving a string of payoffs $Z$ from $j$, and let $Q(Z) = \pi(Z)/\tau$.

By construction, player $i$ will refuse all further game offers from player $j$ if and only if $Q(Z)$ is less than 1, implying that $i$ finds $j$ intolerable. Given any $Z$, the $\tau$, $\pi_0$, and $\omega$ parameter space can then be partitioned into intolerance and tolerance regions characterized by $Q(Z) < 1$ and $Q(Z) \geq 1$, respectively.

Our simulation findings regarding the emergence of wallflower ecologies in the case $\pi_0 < 3.0$ can now be summarized as follows. Using $\{L, D, C, H\}$ to denote the PD game payoffs $\{0, 1, 3, 5\}$, wallflower ecologies rarely evolve in simulations for which the IPD/CR parameters satisfy $\pi_0 < C$, $Q(L) < 1$, $Q(CL) < 1$, and $Q(CCL) \geq 1$. However, we observe at least some ecologies evolving into wallflower ecologies when $\pi_0 < C$ and $Q(CCL) < 1$. For example, given $\omega = 0.6$ and $\tau = W = 1.6$, the value $Q(CCL) = 1$ occurs at $\pi_0 = 2.07$. In simulation experiments with $\omega = 0.6$ and $\tau = W = 1.6$, wallflower ecologies emerged when $\pi_0$ was set at 1.90 or 2.05 but not when $\pi_0$ was set at 2.10 or 2.20.

We also hypothesize that the probability an ecology will evolve into a wallflower ecology depends in part on the initial player population. In particular, suppose parameter values are set so that players in the initial population that do not mutually cooperate are rather quickly reduced to wallflowers and any single cooperator cannot out-score defectors in an otherwise defecting population. Then defectors will take over unless there are two or more cooperators in the initial population; for any time a single cooperator appears via mutation and crossover, it will immediately fail to propagate. On the other hand, if a single cooperator can out-score an otherwise defecting initial population, then cooperators will eventually have offspring with whom they can cooperate, do even better, and take over.

Interestingly, it follows from Table 2 that initial populations whose players have sixteen-state IPD machine representations include on average only 1.47 players whose self-play string is purely cooperative, whereas populations whose players have one-state IPD machine representations have a much greater expected number of self-play cooperators, 7.5. The probability that an initial population will have more than one player that consistently cooperates with large numbers of other players is thus much greater in the one-state case. It follows from the previous discussion that the evolution of wallflower ecologies should be a more common occurrence in our sixteen-state simulations than in our one-state simulations, and this is certainly supported by our simulation findings. As will be discussed further below, *no* wallflower ecologies have been observed in our one-state simulations.

It also follows that wallflower ecologies will be less likely to emerge when the population size is increased. This hypothesis was tested by choosing two of the parameter settings where wallflower ecologies were seen, but incentives were nonexistent: namely, $\omega = 0.3$ and $\omega = 0.4$ with other parameters at standard values. The population size was doubled to $N = 60$, giving an expected number of 2.94 self-cooperators in the initial population. The number of iterations was increased to 400 to ensure that all players had a chance to test all other players. As expected, no wallflower ecologies were seen.

### 5.2.3 The Intermediate Fitness Region $D^1$

In our one-parameter and two-parameter sensitivity experiments, the $D^1$ region is absent for low or intermediate values of the initial expected payoff, $\pi_0$, or for parameter settings where mutual defection results in immediate play stoppage—for example, where $\tau$ is at least 2.4 and all other parameters are set at standard scenario levels. Otherwise, at least some hint of the region is evident, even when the number $I$ of iterations in the standard scenario is increased to 400 so that the average fitness anchoring $D^1$ is close to the average fitness 3.0 that (always) anchors $D^0$.

The $D^1$ region encompasses a whole host of ecologies in addition to the canonical case where players mutually defect once but otherwise mutually cooperate. Many $D^1$ ecologies consist of two or more distinct, repeatedly-interacting subpopulations that appear to be meta-stable in the following sense: they emerge and persist for many generations, and they resist invasion by a large range of mutants.

The social networks formed by the populations inhabiting the $D^1$ ecologies were studied by considering the players as nodes in a fully connected graph. An edge in the graph represented an interaction between two players, and was initially assigned a zero strength. The strength of an edge connecting any two players increased by one each time the players engaged in a PD game. To obtain a clearer visualization of the more persistent social interactions, edge strengths that fell below a certain minimum number at the end of a simulation run were set to zero and the remaining edge strengths were set to one.

Our simulation studies show that the social networks (graphs) determined in this manner for $D^1$ ecologies can take on an astonishing variety of forms: small, isolated networks; linked stars, where a few cooperative players are chosen by all other players; unlinked stars, where the players forming the center nodes of the stars find each other intolerable; a completely connected central network of players together with an outer layer of players, each linked to only one player in the central network; and tree-shaped networks. In many cases, the first ten to fifty iterations are used by players to "get to know each other," after which the social network settles into a seemingly stable configuration. In other cases, however, a social network may persist in one form for a number of generations and then transit to a different form, or it may simply never settle down. New social networks are still being discovered.

As these findings suggest, the $D^1$ region tends to be thick: most of the ecologies in this region do not have an average fitness at the canonical $D^1$ value, just one close to this value. Indeed, one of the ecologies that commonly adds mass and visual impact to the $D^1$ region, Raquel-and-the- Bobs, has a fitness trace that leaps out of the region at random intervals to slightly above 3.0 and then falls back. See, for example, Figure 3. A more detailed study of this and other social networks arising in the evolutionary IPD/CR can be found in Smucker et al. (1994), but a general description of Raquel-and-the-Bobs will now be given.

This whimsically-named ecology is best described as the intertwining of two interaction patterns repeatedly arising from one another. The first interaction pattern occurs when an essentially homogeneous population of "Bobs" evolves. Each Bob has a self-play string containing a single defection at or near the beginning of the self play string, e.g. *d*:*c*, and the property that a point mutation or crossover can easily produce a "Raquel" which is self-cooperative and which also cooperates at all times with a Bob. The population of Bobs obtains an average fitness in the $D^1$ region. A population of Bobs will play with a typical search and latch pattern – a Bob resents the initial defection and keeps searching until it has played all other players once. After that point, it will latch onto the player it has played the largest number of times.

The Bobs hum along from generation to generation until one of their offspring is a Raquel, ushering in the second interaction pattern. The payoff of 5 that a Bob receives from a Raquel when Bob defects and Raquel cooperates causes each Bob to latch onto Raquel as soon as it finds it.

The result of this is that Raquel plays a large number of cooperative games after the initial defect with every Bob in the population and ends up, due to unavoidable stochastic Bob-on-Bob defections in the initial play, with the highest fitness. The population average fitness also rises – not all possible Bob-on-Bob defections happen anymore – and a spike in the fitness trace starts.

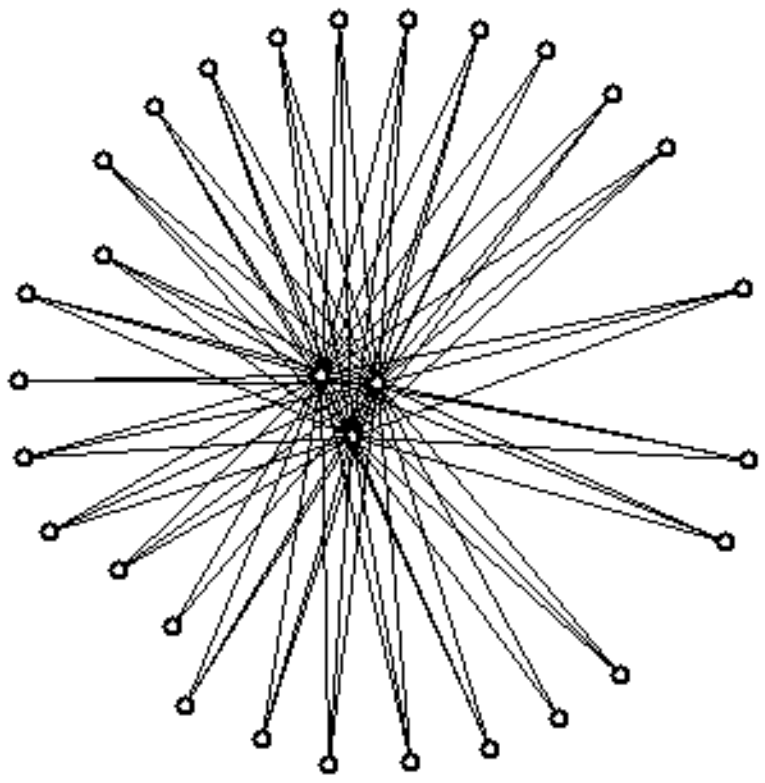

Figure 8: An illustration of a Raquel-and-the-Bobs social network with three Raquels.

Raquel has the highest fitness, so it has children. Since the expressed part of Raquel's genome is usually quite small, the probability is high that Raquel's offspring will also be Raquels. As a result, the number of Raquels increases. With multiple Raquels, a Bob finds a Raquel sooner and gets fewer defection payoffs from other Bobs. Even though a Bob wants to latch onto the first Raquel it finds, the Raquels keep searching the population (at least when $\pi_0 \geq 3$). Thus each Bob eventually finds all the Raquels and obtains the payoff of 5 once for each Raquel. As a result, the population average fitness is raised again. Once enough Raquels have entered the population, the Bobs' fitnesses surpass Raquels'. At this point, given our elitism in reproduction, the Raquels are always decimated and usually wiped out. This causes the falling leg of the spike in the average fitness trace. Figure 8 shows the interaction of a population when there are three Raquels.

#### 5.2.4 Other Phenomena

Reviewing Tables 4 and 5, it is apparent that the fitness region $D^2$ is much less in evidence than the wallflower region or the regions $D^0$ and $D^1$. Typically, only one or two ecologies persist in this region even when it is observed. As for $D^1$, the $D^2$ ecologies that have been observed exhibit a wide variety of behaviors. The fitness region $D^3$ appears to be even rarer than $D^2$, perhaps because three defections often leads to play stoppages. Cyclic ecologies, in which all or almost all individuals have a short repeating self-play string with both $c$'s and $d$'s (e.g. $c : dc$) also occur, but seem to be rare at any parameter setting.

When the standard scenario was perturbed by lowering $\omega$ to 0.5, a particularly interesting ecology appeared whose average payoff persisted near 2.0 over the final twenty generations. Examining the player population for the very last generation, we discovered that the players were alternating mutual defections and mutual cooperations with each other, resulting in sequences of payoffs of the form $(..., 1, 3, 1, 3, ...)$. This population contained two subpopulations, each of whose members preferred partners from the other group.

Sometimes an ecology will transit to a wallflower ecology after persisting for some time in a more cooperative mode. As indicated in Table 4, these rare late wallflower ecologies occur at different parameter values than the usual wallflower ecologies, for reasons that remain unclear.

## 5.3 Sensitivity to Potential Behavioral Complexity

To test the sensitivity of our evolutionary outcomes to changes in the maximum permitted complexity of the players' IPD strategies, the number of states in the IPD machines for these strategies was reduced from sixteen to one. To our surprise, many of the novel behaviors resulting from the introduction of choice and refusal in the evolutionary IPD are not affected by this apparently severe constraint. For one-state IPD machines, $D^2$ ecologies cannot arise; and wallflower ecologies do not emerge at high $\tau$ values with $W$ set equal to $\tau$, apparently because the fraction of cooperators in the initial population is high; cf. Table 2. However, $D^1$ ecologies (in particular, Raquel-and-the-Bobs) are possible and have been observed.

## 5.4 Floating $\tau$ and $\omega$ Studies

In some studies, rather than assuming that the players were characterized by fixed commonly-shared values for the minimum tolerance level $\tau$ and the memory weight $\omega$, we instead let these parameters constitute part of each player's genetic structure. For these simulations, we used the Moore machine representation for the players' IPD strategies. Sixteen bits for each of the two parameters were added to the bit strings used to code the Moore machine representations. These additional 16 bits were used to partition the allowable ranges for $\tau$ and $\omega$ into $2^{16}$ intervals. For the floating $\tau$ study, the allowable range was set from 0 to 3; for the floating $\omega$ study, the allowable range was set from 0 to 1.

Figure 9 shows what happened when $\tau$ was added to the genetic structure of each player, with all other parameters set at standard scenario levels (including $W = 1.6$). We made 196 runs from different initial random seeds, with each run consisting of 2000 generations. The average $\tau$ across the 196 runs evolved to approximately 2.1. [Recall that $\tau = 2.1$ is the value above which any sucker payoff 0 results in further refusal of play.] Moreover, for all but the initial generations, the overall average fitness across the 196 runs hovered around 2.8. Surprisingly, this is generally lower than the overall average fitness achieved in the standard scenario case with $\tau$ held fixed at $W = 1.6$; see part (a) of Figure 3.

When the ecologies were examined individually, however, we found once again a rich variety of behaviors. The average values of $\tau$ for individual ecologies have an interesting distribution (see Fig. 9(b)). Average $\tau$ values tend to cluster in two intervals at later generations, (2.1-2.25) and (2.7-3.0). A third peak is observed around $\tau = 1.4$, and there is a gap between this lower region and $\tau = 1.7$ which is uninhabited by any ecologies. Figure 9(c) highlights the appearance of wallflower ecologies at high $\tau$ values. These wallflower ecologies help account for the lower average fitness achieved with floating $\tau$ compared to the standard scenario with fixed $\tau$. Thus a high $\tau$ carries a significant risk.

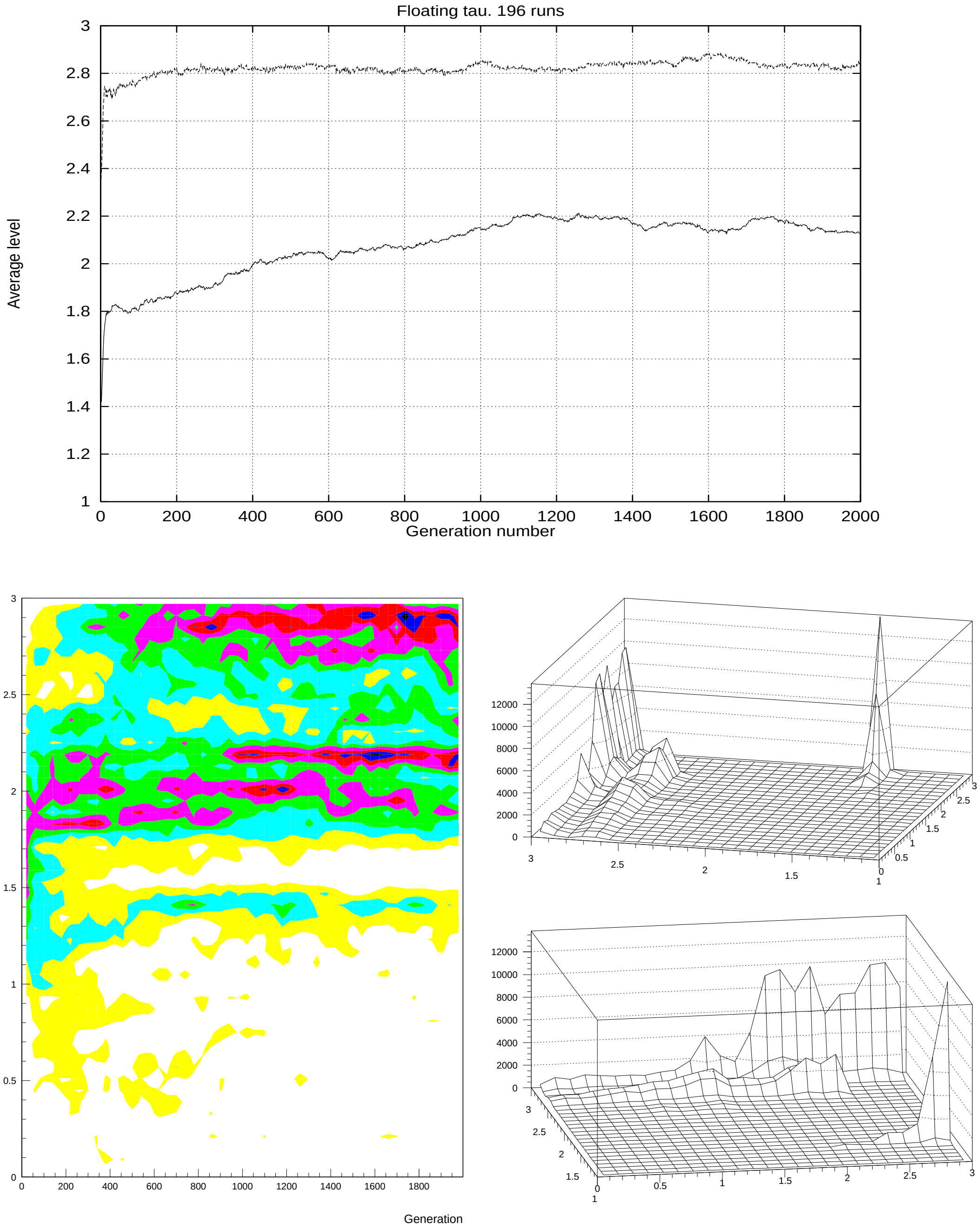

Figure 9: Floating $\tau$ results. (a) The average fitness and average $\tau$ across 196 runs for 2000 successive generations. (b) The distribution of the mean $\tau$ for each generation. In this contour plot, levels are shaded from white at low density levels to dark at high levels. (c) The joint distribution of the mean average fitness and the mean $\tau$, at two different angles. Mean average fitness is plotted from 1 to 3 and mean $\tau$ from 0 to 3.

When the above experiment was repeated with a floating $\omega$ and with $\tau$ fixed at its standard scenario level 1.6, the floating $\omega$ evolved on average to a very low level. As for the floating $\tau$ case, the mean average fitness reached was lower than for the standard scenario case, and this mean masked a rich variety of behaviors. These results are shown in Figure 10. Interestingly, the mean value of $\omega$ never rose above 0.53, the value where $Q(L) = 1$. However, the distribution of values against generation shows that many ecologies did in fact have high average $\omega$. Low $\omega$ is associated with a large number of wallflower ecologies. Note that the regions of high and low distribution values in the floating $\omega$ contour plot of Figure 10(c) are much less distinct than the analogous ones in the floating $\tau$ results. Figure 10(d) once again shows the danger of being too intolerant of defection, for wallflower ecologies tend to be limited to the range (0-0.3).

### 5.5 Population Diversity

Our small populations quickly tend to lose genetic diversity. To a certain extent, this is an inherent property of small populations, but to a certain extent it is also an artifact of the way in which we have implemented our genetic algorithm. If we increased the mutation rate, decreased the number of elite, or replaced elitism with reproduction proportional to fitness, then presumably the fitness regions would become noisier and convergence to a stable level of genetic diversity less likely. An effect that we did not anticipate, however, is that the behavioral diversity of our populations is sensitive to the choice of sorting algorithm used to select parents, even though the genetic diversity is not.

More precisely, in most of our simulations we use a bubble-sort algorithm to rank potential parents by fitness. The bubble-sort algorithm is biased towards the incumbency of older players. That is, in cases where a subset of players all have the same fitness, as often occurs, the players that achieved relatively higher rankings at the end of the previous generation are again ranked higher, making their survival more likely. This incumbency effect, combined with our use of strong elitism ($X = 20$), tends to promote the evolution of behaviorally homogeneous populations.

Use of a heap-sort algorithm gives slightly noisier results but in general does not appear to affect our conclusions. Under a randomized sorting algorithm, however, the fitness regions are noisier yet, and, although many of the features derived using bubble-sort remain, they are less prominent. For example, Raquel-and-the-Bobs ecologies occur much less frequently and regularly.

## 6 Concluding Remarks

The simulation studies reported in the present paper, in the companion paper Smucker et al. (1994), and in Stanley et al. (1994) indicate that permitting players in an evolutionary IPD to choose and refuse potential partners can have a significant effect on the results. A key issue that remains to be explored, however, is whether the effects we have found are generic to choice and refusal or are closely tied to particular features of our implementation.

For example, our reliance on a synchronized genetic step, strong elitism, expected payoffs, and a strict tolerance threshold may affect our outcomes significantly. Also, using a fixed convex combination of past expected payoff and current payoff in order to obtain an updated expected payoff is too rigid and simplistic, even in the present context in which players have no prior knowledge of other players' strategies and payoffs. Allowing the minimum tolerance level $\tau$ and the memory weight $\omega$ to evolve is a step towards increased flexibility. More generally, however, what would happen if players were free to evolve their partner selection mechanisms from a broader domain? Some players might exhibit a preference for those they have played before whereas others might display a taste for variety. Also, we might see the emergence of a more sophisticated form of anticipatory behavior in the form of signals among potential partners meant to influence future partner selection.

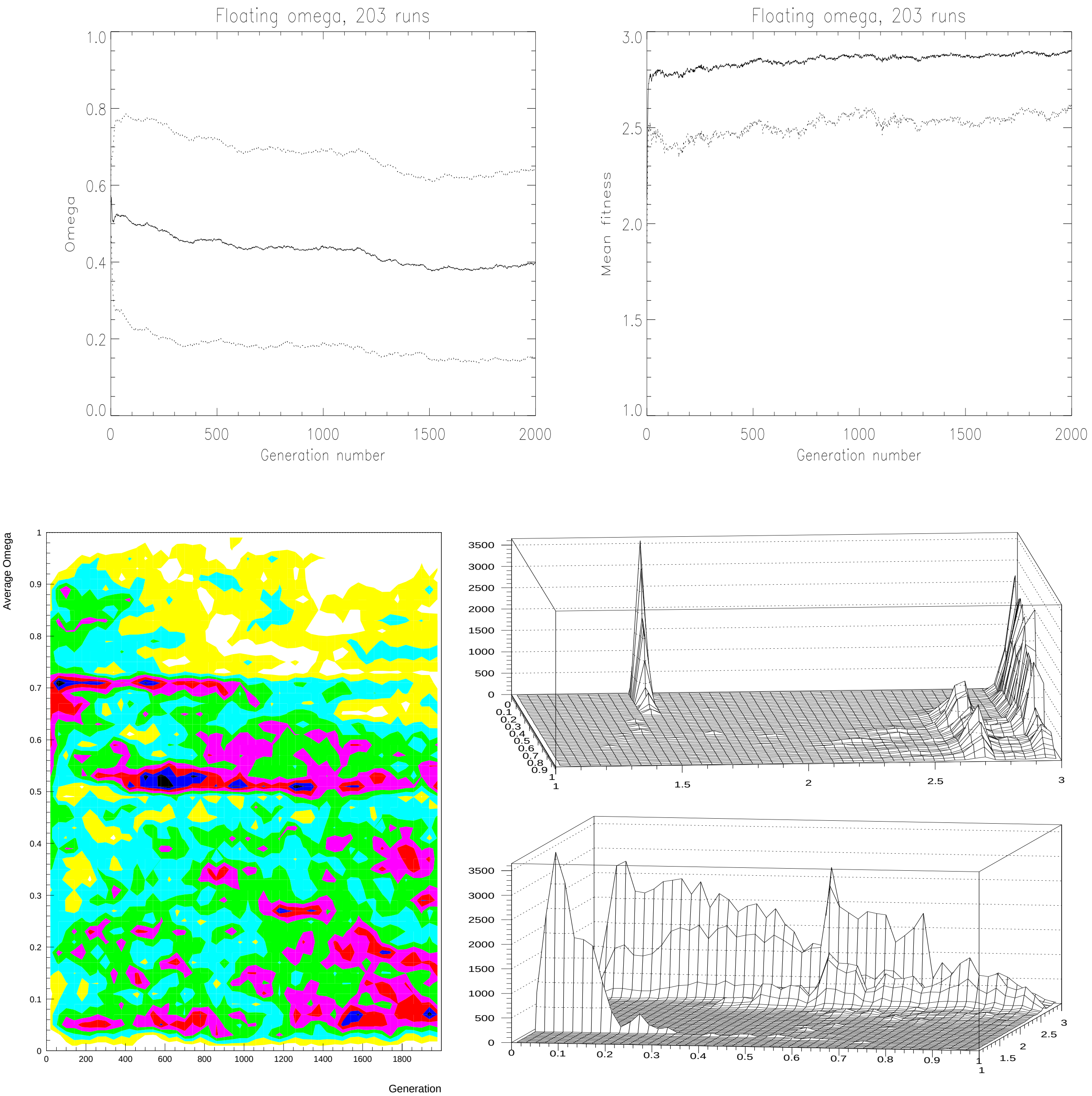

Figure 10: Floating $\omega$ results. (a) The mean $\omega$ across 203 runs for 2000 successive generations, together with error bounds (plus or minus one standard deviation). (b) The mean average fitness across 203 runs for each generation, with a lower error bound. (c) The distribution of the mean $\omega$ for each generation. In this contour plot, levels are shaded from white at low density levels to dark at high levels. (d) The joint distribution of the mean average fitness and the mean $\omega$, at two different angles. Mean average fitness is plotted from 1 to 3 and mean $\omega$ is plotted from 0 to 1.

Our long-term interest is to develop realistic models of human interactions in social and economic contexts. The addition of choice and refusal of partners using expected payoffs is intended to be a step in that direction. However, the PD game is a caricature of social interactions. Applying the insights developed in the present paper to real situations will require the modeling of more specific cases. One of our original goals was to develop a model of sexual partner selection that could help in understanding and controlling the spread of AIDS. Another possible application is to the endogenous formation of economic institutions; see, for example, Holland (1992, Chapter 10) and Marimon et al. (1992). Tesfatsion (1994) has modified the IPD/CR framework to develop a trade-coalition game with preferential selection of trading partners and evolved trading networks.

## Acknowledgments

This research was partially supported by an Iowa State University Research Grant funded under DHHS Grant # 2SO7RR07034-26 and by the Los Alamos Center for Nonlinear Studies. Many thanks to the Iowa State Physics and Astronomy Department, especially the Gamma Ray Astronomy research group, for providing computer facilities and office space to Mark Smucker.